\documentclass[11pt,draftcls,onecolumn,journal]{IEEEtran}
\usepackage{algorithm}
\usepackage{algorithmic}
\usepackage{amsmath,amsfonts}
\usepackage{array}
\usepackage{algorithmic}
\usepackage{xcolor}
\usepackage{algorithm}
\usepackage{array}
\usepackage[caption=false,font=normalsize,labelfont=sf,textfont=sf]{subfig}
\usepackage{textcomp}
\usepackage{stfloats}
\usepackage{url}
\usepackage{verbatim}
\usepackage{graphicx}
\usepackage{cite}

\usepackage[overload]{empheq} 

\usepackage{tikz}
\usetikzlibrary{arrows}

\usepackage[most]{tcolorbox}
\newtcolorbox[auto counter]{mybox}[2][]{%
    title=Box \thetcbcounter: #2, #1}

\begin{document}
%
\title{Fast Computation of the Discrete Fourier Transform Square Index Coefficients}
\author{Saulo Queiroz, 
João P. Vilela, and Edmundo Monteiro, \IEEEmembership{Senior IEEE}
\thanks{Saulo Queiroz (sauloqueiroz@utfpr.edu.br) is with the Academic Department of Informatics of the Federal University of Technology Paran\'a (UTFPR), Ponta Grossa, PR, Brazil, the Centre for Informatics and Systems of the University of Coimbra (CISUC) and the Department of Computer Science of the University of Porto, Porto, Portugal.
}
\thanks{Jo\~ao P. Vilela (jvilela@fc.up.pt) is with CRACS/INESCTEC, CISUC and the Department of Computer Science, Faculty of Sciences, University of Porto, Portugal.}
\thanks{Edmundo Monteiro (edmundo@dei.uc.pt) is with the Department of Informatics Engineering of the University of Coimbra and CISUC, Portugal.}
}

%
\markboth{Copyright transferred to IEEE for publication in IEEE Signal Processing Magazine.}%
{Queiroz \MakeLowercase{\textit{et al.}}: Fast Computation of Square Index DFT Coefficients}


\maketitle
%
%
%
%
%
%
%
\IEEEPARstart{T}{he} $N$-point discrete Fourier transform (DFT)
 is a cornerstone for several signal processing applications. 
Many of these applications operate in real-time, making the 
computational complexity of the DFT a critical performance indicator 
to be optimized. Unfortunately, whether the $\mathcal{O}(N\log_2 N)$ time complexity of 
the fast Fourier transform (FFT) can be outperformed remains an unresolved 
question in the theory of computation. However, in many applications of the 
DFT -- such as compressive sensing, image processing, and wideband spectral 
analysis \cite{sfft-survey-2014} -- only a small fraction of the output signal needs 
to be computed because the signal is sparse. This motivates the development of 
algorithms that compute specific DFT coefficients more efficiently than the FFT
 algorithm. In this article, we show that the number of points of some DFT 
coefficients can be dramatically reduced by means of elementary mathematical properties. 
We present an algorithm that compresses the square index coefficients (SICs) of 
DFT  (i.e., $X_{k\sqrt{N}}$, $k=0,1,\cdots, \sqrt{N}-1$, for a square number $N$) 
from $N$ to $\sqrt{N}$ points at the expense of $N-1$ complex 
sums and no multiplication. Based on this, any regular DFT algorithm 
can be straightforwardly applied to compute the SICs with a reduced number
of complex multiplications. If $N$ is a power of two, one
can combine our algorithm with the FFT algorithm to calculate all SICs in
$\mathcal{O}(\sqrt{N}\log_2\sqrt{N})$ time complexity.

\section*{Compressing Square Index DFT Coefficients}
Consider the classic DFT computation $X_k$ ($k=0,1,\cdots,N-1$) of the input signal $x_n$ ($n=0,1,\cdots,N-1$) 
in which the complex exponential $e^{j2\pi/N}$ is denoted by $W_N$, i.e.,
\begin{eqnarray}
X_{k} &=& \sum_{n=0}^{N-1}x_nW_N^{-kn}. \label{eqn:dft}
\end{eqnarray}

In what follows, we will demonstrate that the number of points in (\ref{eqn:dft})
can be reduced from $N$ to $\sqrt{N}$ for the SICs at the expense of $N-1$ complex sums. 
As we will see later, these DFT coefficients have several practical applications.

Let us assume $N$ is a perfect square number.
In this case, the input
sample $x_n$ can be placed at the $l$-th row and $c$-th column of a 
$\sqrt{N}\times\sqrt{N}$ square arrangement such that 
\begin{eqnarray}
n&=&l\sqrt{N}+c \label{eqn:n},
\end{eqnarray}
for $l=0,1,\cdots,\sqrt{N}-1$ and $c=0,1,\cdots,\sqrt{N}-1$. 
This idea is illustrated next for the computation of the 
coefficient $X_0$ considering $N=16$ (i.e., $W_{16}^{-0\cdot n}=1$).
\begin{eqnarray}
X_{0}  &=& x_{0 \sqrt{16}+0} + x_{0\sqrt{16}+1} + x_{0 \sqrt{16}+2} + x_{0 \sqrt{16}+3} +  \nonumber \\
       & & x_{1 \sqrt{16}+0} + x_{1 \sqrt{16}+1} + x_{1 \sqrt{16}+2} + x_{1 \sqrt{16}+3} + \nonumber \\
       & & x_{2 \sqrt{16}+0} + x_{2 \sqrt{16}+1} + x_{2 \sqrt{16}+2} + x_{2 \sqrt{16}+3} + \nonumber \\
       & & x_{3 \sqrt{16}+0} + x_{3 \sqrt{16}+1} + x_{3 \sqrt{16}+2} + x_{3 \sqrt{16}+3}.
\label{eqn:step1}  
\end{eqnarray}
By applying (\ref{eqn:n}) to (\ref{eqn:dft}) one gets
\begin{eqnarray}
X_{k} &=& \left(\sum_{c=0}^{\sqrt{N}-1}\sum_{l=0}^{\sqrt{N}-1}x_{l\sqrt{N}+c}W_N^{-k(l\sqrt{N}+c)}\right). \label{eqn:v1} 
\end{eqnarray}

Note that the positions of the summation symbols in (\ref{eqn:v1}) are interchangeable due 
to the commutative property of summations. The order we choose is for the convenience of our 
proof next.
 
Let us consider the case of SICs, i.e., output coefficients $X_{k\sqrt{N}}$, 
$k=0,1,\cdots,\sqrt{N}-1$. In this case, Eq.~(\ref{eqn:v1}) rewrites as, 
\begin{eqnarray}
X_{k\sqrt{N}} &=& \left(\sum_{c=0}^{\sqrt{N}-1}\sum_{l=0}^{\sqrt{N}-1}x_{l\sqrt{N}+c}W_N^{-k\sqrt{N}(l\sqrt{N}+c)}\right). \label{eqn:v2} 
\end{eqnarray}
Since $W_N^{\sqrt{N}^2}$ results in a root of unity (i.e., $W_N^{-kl\sqrt{N}^2}=1$), 
and $W_N^{-k\sqrt{N}c}=W_{\sqrt{N}}^{-kc}$,
(\ref{eqn:v2}) simplifies to
\begin{eqnarray}
X_{k\sqrt{N}} &=& \left(\sum_{c=0}^{\sqrt{N}-1}\sum_{l=0}^{\sqrt{N}-1}x_{l\sqrt{N}+c}W_{\sqrt{N}}^{-kc}\right).
\label{eqn:v3} 
\end{eqnarray}
Note that the complex exponential in (\ref{eqn:v3}) is independent of $l$. Based
on this, it results
\begin{eqnarray}
X_{k\sqrt{N}} &=& \left(\sum_{c=0}^{\sqrt{N}-1}W_{\sqrt{N}}^{-kc}\sum_{l=0}^{\sqrt{N}-1}x_{l\sqrt{N}+c}\right).
\label{eqn:v4} 
\end{eqnarray}
By denoting the inner summation of (\ref{eqn:v4}) as
\begin{eqnarray}
\hat{x}_c &=& \sum_{l=0}^{\sqrt{N}-1}x_{l \sqrt{N}+c}\label{eqn:xhat},
\end{eqnarray}
Eq.~(\ref{eqn:v4}) rewrites to
\begin{eqnarray}
X_{k\sqrt{N}} &=& \sum_{c=0}^{\sqrt{N}-1}W_{\sqrt{N}}^{{- k c}}{\hat{x}_c},  \label{eqn:compacdft}
\end{eqnarray}

Note that (\ref{eqn:compacdft}) is a compressed version of the original DFT (\ref{eqn:dft}) 
for SICs. In other words, by performing (\ref{eqn:xhat}) (for $c=0,1,\cdots,\sqrt{N}$) 
on the $N$-point input signal array $\mathbf{x}=\{x_0,\cdots,x_{N-1}\}$, one gets 
the compressed $\sqrt{N}$-point input signal array 
$\mathbf{\hat{x}}=\{\hat{x}_0,\hat{x}_1,\cdots,\hat{x}_{\sqrt{N}-1}\}$
at the computational cost of only $N-1$ complex sums. To compute the output array 
of coefficients $\mathbf{\hat{X}}=\{\hat{X}_0,\hat{X}_1,\cdots,\hat{X}_{\sqrt{N}-1}\}$, 
a $\sqrt{N}$-point DFT on $\mathbf{\hat{x}}$ will vary $k$ from $0$ to $\sqrt{N}-1$.
Thus, the obtained coefficients match the $\sqrt{N}$ SICs of the original input $\mathbf{x}$
following the correspondence  $X_{k\sqrt{N}}= \hat{X}_{k}$. 
Next, we exemplify how to perform the DFT of SICs based on
the compressed DFT signal (\ref{eqn:compacdft}).

\section*{Numerical Example}
Consider the following example of a $N=9$-point signal,
$$\mathbf{x} =\{11+11j, 22+22j, 33+33j, 
      -5-5j, -6-6j, -7-7j, 
       9-9j, 10-10j, 11-11j \}.$$
The first step consisting in computing the multiplierless summation 
(\ref{eqn:xhat}). It results from adding the samples of 
$\mathbf{x}$ at every $\sqrt{9}=3$ step to get the smaller
vector $\mathbf{\hat{x}} =\{\hat{x}_0, \hat{x}_1, \hat{x}_2 \}$. This yields,
\begin{eqnarray}
\hat{x}_0&=&\sum_{l=0}^{\sqrt{9}-1}x_{l\sqrt{9}+0}=11+11j-5-5j+9-9j=15-3j,\\
\hat{x}_1&=&\sum_{l=0}^{\sqrt{9}-1}x_{l\sqrt{9}+1}=22+22j-6-6j+10-10j=26+6j,\\
\hat{x}_2&=&\sum_{l=0}^{\sqrt{9}-1}x_{l\sqrt{9}+2}=33+33j-7-7j+11-11j=37+15j.
\end{eqnarray}
A DFT on $\mathbf{\hat{x}}$ will produce the output signal 
vector $\mathbf{\hat{X}}=\{\hat{X}_0,\hat{X}_1,\hat{X}_2\}$. 
Note that $X_{k\sqrt{N}}=\hat{X}_k$, as we mentioned before.
Therefore, by performing a DFT on $\mathbf{\hat{x}}$, one gets
\begin{eqnarray}
X_{0\sqrt{9}} &=& \hat{X}_0  = \sum_{c=0}^{\sqrt{9}-1}W_{\sqrt{9}}^{{- 0\cdot c}}{\hat{x}_c}\approx 78 + 18j,  \\
X_{1\sqrt{9}} &=& \hat{X}_1  = \sum_{c=0}^{\sqrt{9}-1}W_{\sqrt{9}}^{{- 1\cdot c}}{\hat{x}_c}\approx -24.2942 -3.9737j,  \\
X_{2\sqrt{9}} &=& \hat{X}_2  = \sum_{c=0}^{\sqrt{9}-1}W_{\sqrt{9}}^{{- 2\cdot c}}{\hat{x}_c}\approx -8.7058 - 23.0263j.
\end{eqnarray}

\begin{figure*}[t]
\centering
  \includegraphics[scale=1]{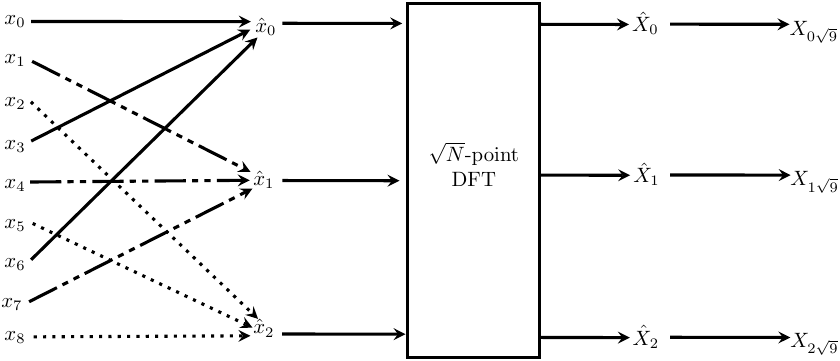}
  \caption{Butterfly diagram for the computation of the $9$-point DFT coefficients
$X_{0\sqrt{9}}$, $X_{1\sqrt{9}}$, and $X_{2\sqrt{9}}$ of the input 
signal ${\mathbf{x}}=\{x_0,\cdots,x_8\}$.
Firstly, ${\mathbf{x}}$ is compressed into the signal $\mathbf{\hat{x}}=\{\hat{x}_0, \hat{x}_1, \hat{x}_2\}$ 
according to (\ref{eqn:xhat}). Then, a $\sqrt{9}$-point DFT on $\mathbf{\hat{x}}$ 
results in the coefficients $\hat{\mathbf{X}}=\{\hat{X}_0,\hat{X}_1,\hat{X}_2\}$ 
such that $X_{k\sqrt{9}}=\hat{X}_k$ ($k=0,1,2$).
}
\label{fig:butterf}
\end{figure*}

Fig.~\ref{fig:butterf} illustrates the above steps in a summarized
form based on the well-known butterfly diagram of DFTs. 
Note that getting $\{\hat{x}_0,\hat{x}_1,\hat{x}_2\}$ from $\{x_0,\cdots,x_8\}$ 
based on (\ref{eqn:xhat}) dispenses the FFT's twiddle factors. 
A regular DFT implementation is employed to compute
the DFT $\mathbf{\hat{X}}$ of the $\sqrt{N}$-point compressed signal 
$\mathbf{\hat{x}}$. 

For inverse FFT (IFFT) implementations that apply
a normalization factor $K$ to the inverse DFT (IDFT) coefficients, 
it is important to consider the impact of the signal length on $K$. For example, if 
an IFFT implementation normalizes the coefficients by the reciprocal of the 
signal length  (i.e., $K=1/N$), then an input signal 
of length  $\sqrt{N}$ will be normalized by $K=1/\sqrt{N}$. Therefore, 
by performing an FFT on the $\sqrt{N}$-point signal compressed by our algorithm 
(instead of on the original $N$-point signal), the resulting DFT coefficients
will be normalized by $K=1/\sqrt{N}$ rather of $K=1/N$, as required. To address this,
one can multiply the obtained IDFT coefficients by an additional normalization factor
$\hat{K}=1/\sqrt{N}$ so to achieve the desired final normalization of $1/N$,
since $\hat{K}K=(1/\sqrt{N})(1/\sqrt{N})=1/N$. Similarly, 
if normalization $K=1/\sqrt{N}$ is considered by an FFT implementation to 
comply the Parseval's theorem and preserve the signal energy after the transform,
then $\hat{K}$ should be set to $\sqrt{\frac{1}{\sqrt{N}}}$, so that
$\hat{K}K=(1/\sqrt{\sqrt{N}})(1/\sqrt{\sqrt{N}})=1/\sqrt{N}$, preserving the signal
energy.

\begin{algorithm}[t]
\caption{SIC DFT Compression Algorithm.\label{alg:compactsic}}
\begin{algorithmic}[1]
\STATE \textbf{Input:} $\mathbf{x}$ (vector), $N$ (length of $\mathbf{x}$)
\STATE \text{allocate the vector} $\hat{\mathbf{x}}[0,\cdots,\sqrt{N}-1]\gets 0$;
\FOR{$c = 0$ \TO $\sqrt{N} - 1$}
    \STATE $\hat{\mathbf{x}}[c] \gets 0$;
    \FOR{$l = 0$ \TO $\sqrt{N} - 1$}
        \STATE $\hat{\mathbf{x}}[c] \gets \hat{\mathbf{x}}[c] + \mathbf{x}[c + l \sqrt{N}]$;
    \ENDFOR
\ENDFOR
\RETURN $\hat{\mathbf x}$;
\end{algorithmic}
\end{algorithm}

\section*{The Trick for Fast Computation of DFT SICs}
In Alg.~\ref{alg:compactsic}, we present the SIC DFT
compression algorithm (\ref{eqn:compacdft}). The algorithm takes an $N$-point signal
$\mathbf{x}$ as input and gives its corresponding $\sqrt{N}$-point 
compressed signal $\mathbf{\hat{x}}$ as output. 
The trick consists in preceding a regular DFT implementation with 
Alg.~\ref{alg:compactsic} to achieve faster computation of SICs.
Note that the algorithm performs no complex multiplication and consists only of $N-1$ 
complex additions. Indeed, the number of iterations in each loop of the algorithm 
is $\sqrt{N}$, therefore its overall asymptotic complexity 
$\mathcal{O}(\sqrt{N})\cdot \mathcal{O}(\sqrt{N})=\mathcal{O}(N)$ complex additions. 
After employing our algorithm, any regular DFT implementation can be used for the calculation of
 the output coefficients.
If the DFT implementation has a complexity of $T(N)$, then preceding it with
our algorithm will result in a complexity of $T(\sqrt{N})$ for the computation
of all SICs.

Consider, for example, the employment of the classic DFT formula (\ref{eqn:dft}).
To compute $N$ coefficients of $N$ points each, one gets a computational 
complexity of $\mathcal{O}(N^2)$ multiplications. If only the $\sqrt{N}$ SICs are desired, 
the resulting asymptotic complexity is 
$\mathcal{O}(\sqrt{N})\cdot \mathcal{O}(N)=\mathcal{O}(N\sqrt{N})$, since
each SIC has $N$ points. With the assistance of our algorithm, the number of
points reduces to $\sqrt{N}$ and the number of multiplications improves to 
$\mathcal{O}(\sqrt{N})\cdot \mathcal{O}(\sqrt{N})=\mathcal{O}(N)$. This provides 
flexibility for the efficient computation of specific frequencies within a spectrum, 
as is often required for harmonic signals. We will demonstrate a practical example of 
this later.

If $N$ is a square power of two, executing an FFT on the signal compressed by our algorithm 
will produce all square index coefficients in $\mathcal{O}(\sqrt{N}\log{\sqrt{N}})$
time complexity. This complexity is optimal if the FFT algorithm 
is proven to be the fastest algorithm for the DFT problem, a question that remains 
open and has implications for important fundamental limits of science, such as the 
DFT lower-bound complexity~\cite{fastenough-2022} and the capacity limits of DFT-based 
communication signals~\cite{queiroz2024complim}.

\section*{Practical Considerations}
The speedup achieved by Algorithm~\ref{alg:compactsic} depends on two 
main conditions: the input signal length $N$ must be a perfect square and 
only DFT coefficients indexed at multiples of $\sqrt{N}$ need to be computed.
Next, we discuss how these conditions can be met to facilitate the 
practical application of our algorithm.

The widespread adoption of the FFT algorithm favours the requisite for
signals of perfect square length. This happens because FFT requires $N$ to
be a power of two and any power of two raised to an even number is also a 
perfect square\footnote{Indeed, since even numbers are of the form
$2m$, where $m=0,1,2,\dots$, raising 2 to the power of $2m$ results in the perfect 
squares with roots given by $\sqrt{2^{2m}}=2^{m}$.}. 
Therefore, by meeting the FFT requirement in these cases, several digital systems also satisfy 
the requirements of our algorithm. This is the case, for instance, with telecommunication 
standards like 5G and IEEE 802.11 (WiFi), that adopt values like 64, 256, 1024, 
and 4096 for certain setups.
For cases where $N$ is not a perfect square, the input signal can be padded with zeros 
so that $N$ becomes a perfect square. This is similar to the common practice of padding 
the FFT input with zeros to meet its power-of-two requirement. 
In our case, the number of zeros required can be calculated as $(\lceil \sqrt{N} \rceil)^2-N$,
where $\lceil \phi \rceil$ denotes the ceiling function that returns the smallest
integer greater than the real number $\phi$.

Another requirement of our algorithm is to compute only frequencies that are multiples
of $\sqrt{N}$. In other words, all frequencies of interest in the observed spectrum must
be multiples of a common value.
A notable example of this is the harmonic signals, which have a broad range
of applications in fields such as telecommunications, acoustics, power transmission,
control theory, etc. 
In a harmonic signal, the frequencies of interest -- known as ``harmonics'' -- 
are all integer multiples of a fundamental frequency $f_0$. Therefore, if $f_0$ is a 
multiple of $\sqrt{N}$ then all other harmonics are also multiples of $\sqrt{N}$, and our algorithm suits.
To accomplish this, one can conveniently adjust either the frequency resolution or the sampling rate.
We illustrate the practical application of our algorithm with a case study in the next section.

\section*{Case Study: Spectral Analysis of the A440 Piano Key}
\begin{figure*}[t]
\centering
  \includegraphics[scale=0.5]{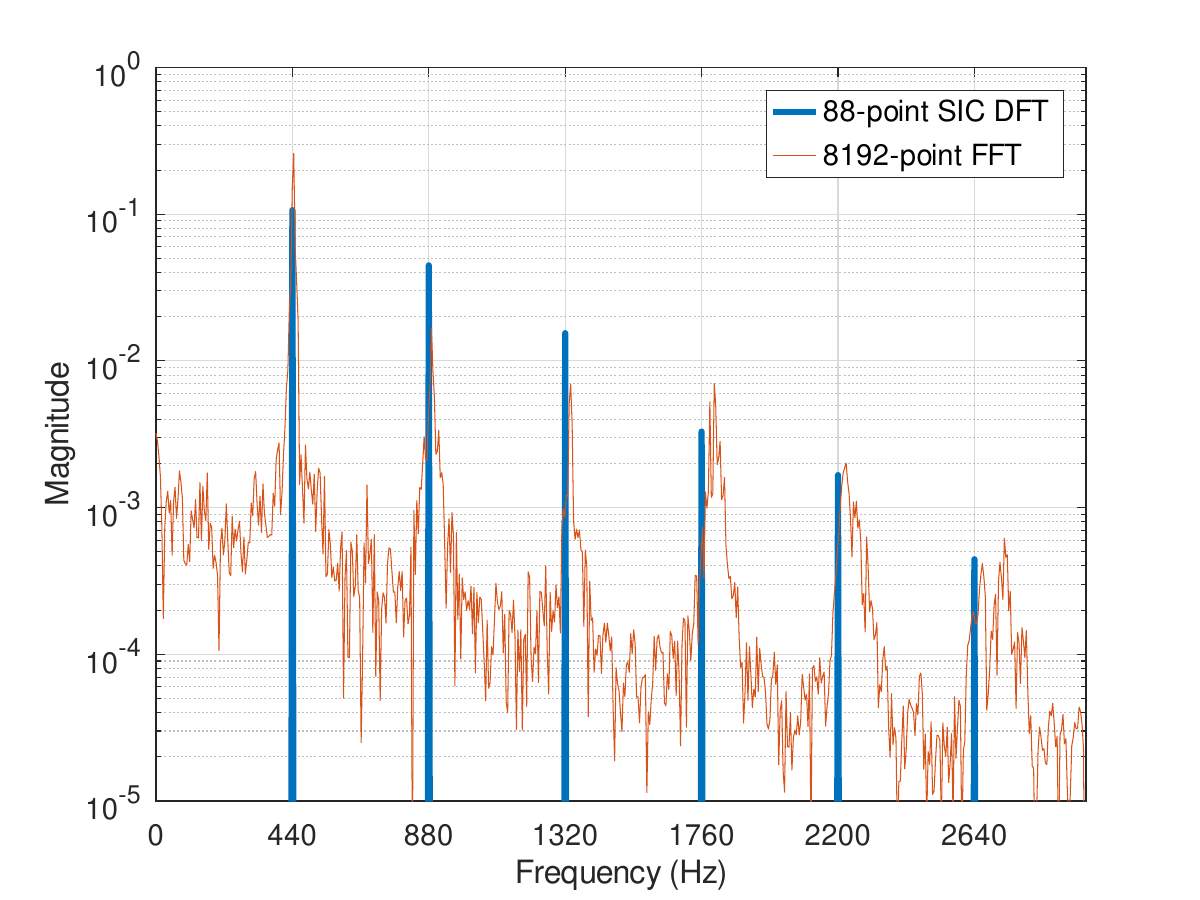}
  \caption{FFT vs. proposed SIC DFT algorithm: 1st to 6th harmonics of the A440 piano key with  fundamental frequency $440$ Hz.}
\label{fig:a4spec}
\end{figure*}

Musical instruments are well-known generators of harmonic frequencies. 
The A440 piano key (a.k.a. ``middle A'', ``concert pitch'', and ``A4''),
for example, has a fundamental frequency  $f_0=440$ Hz and emits fading harmonics 
at frequencies 880 Hz, 1320 Hz, and so on\footnote{Please, refer to the entry 
``A440\_(pitch\_standard)'' on Wikipedia for additional informations.}.  
Fig.~\ref{fig:a4spec} illustrates a
spectral analysis of the A440 piano key up to the 6th harmonic. 
The Matlab script and the audio file from the experiment are publicly available
in the corresponding author's `Github' repository\footnote{https://github.com/sauloqueiroz/fastsicdft/.}.

The red curve in  Fig.~\ref{fig:a4spec} corresponds to an 8192-point FFT on a wave 
file sampled at 38720 Hz, yielding a frequency resolution of $38720/8192\approx 4.72$~Hz.
The blue bars represent the magnitudes computed by our SIC DFT algorithm. 
Despite some unexpected frequencies near the 4th, 5th, and 6th harmonics --
possibly due to characteristics of the piano instrument considered --
our algorithm accurately reveals the higher magnitude frequencies expected 
for an A440 piano key, as shown in the figure. To achieve this,
we conveniently set the input length for our SIC DFT algorithm to the perfect square 
$88^2=7744$. This gives a frequency resolution of $38720/7744=5$ Hz, which is nearly the same 
as the 8192-point FFT.  Additionally, the fundamental frequency of 440 Hz in our experiment is 
a multiple of $\sqrt{7744}=88$, consequently all frequencies of interest are also multiples of 88.

Under this setup, our algorithm calculates the 88-point compressed signal
$\mathbf{\hat{x}}=\{\hat{x}_0,\cdots,\hat{x}_{87}\}$  from the 7744-point input signal
$\mathbf{x}=\{x_0,\cdots,x_{7743}\}$ with a computational cost of 7743 complex additions 
and no complex multiplication. As previously demonstrated, the DFTs of $\mathbf{\hat{x}}$ and $\mathbf{x}$,
denoted as $\mathbf{\hat{X}}=\{\hat{X}_0,\cdots,\hat{X}_{87}\}$ and 
$\mathbf{X}=\{{X}_0,\cdots,{X}_{7743}\}$, respectively, satisfy 
$\hat{X}_k=X_{k\sqrt{7744}}$, for $k=0,\cdots,87$. By executing any DFT algorithm on 
$\mathbf{\hat{x}}$, one can obtain $\mathbf{\hat{X}}$ and thus determine the harmonic 
frequencies of $\mathbf{X}$.
For example, by padding the 88-point input signal with zeros to reach a length of 128 points
(the smallest power of two higher than 88), the FFT algorithm can be used to compute all 88 
frequencies\footnote{Note that the condition for $N$ being a perfect square is a requirement
for the signal to be compressed, not for the already compressed signal.}. 
Note, however, that only six frequencies are relevant for the experiment. 
Thus, one can choose to compute six  88-point frequencies following the regular DFTs algorithm
instead of a 128-point FFT. In this case, approximately $87\cdot 6=522$ complex multiplications
are performed against the $\mathcal{O}(128\log_2128)$ complex multiplications of the FFT algorithm.
Besides, the prime-factor DFT algorithm (PFA) can also be
 employed with complexity $\mathcal{O}(88\log 88)$ in this case, since the input length $88$ 
can be factorized into co-primes (8 and 11) as required by PFA.

In any case, the signal compression achieved by our algorithm offers a significant 
computational performance advantage over the 8192-point FFT when only a specific 
subset of frequencies needs to be observed, as in many scenarios involving harmonic signals.
This characteristic corresponds to very sparse signals. For example, 
considering the frequency resolution of approximately $4.72$~Hz in 
our experiment, more than 600 frequencies are present in the range 
$[0,3000]$ Hz displayed in Fig.~\ref{fig:a4spec}, yet only six of these 
frequencies are of interest. In spite of that, 
sparse FFT algorithms are not suitable in this case because they 
require much larger inputs to outperform FFT, such as $N=2^{18}$ \cite{haitham-soda-2012}.
Therefore, the compression provided by our algorithm serves as a valuable technique to
accelerate the DFT computation of sparse signals.

\section*{Conclusion}
In this article, we demonstrate that the number of points of certain
DFT coefficients can be reduced by means of elementary mathematical
tricks. Leveraging this, any regular DFT algorithm can speed up the
computation of those coefficients by operating on inputs of smaller
sizes. To this end, we present a multiplierless algorithm that performs 
$N-1$ complex additions to compress the number of points 
of SICs from $N$ to $\sqrt{N}$. Furthermore, if $N$ is a power of two, the 
FFT algorithm can be preceded by our algorithm to compute all SICs 
with $\mathcal{O}(\sqrt{N}\log_2 \sqrt{N})$ complex multiplications.
This method can find applications in sparse and pruned DFTs, where only a fraction 
of DFT coefficients are of interest.
Our article poses an interesting question about whether our techniques can inspire new 
methods to speed up the DFT of other patterns of coefficients. In this regard, the 
authors would like to challenge the readers.

\section*{Acknowledgements}
Authors would like to thank The Science and Technology Development Fund, Macau SAR. (File no. 0044/2022/A1) 
and Agenda Mobilizadora Sines Nexus (ref. No. 7113), supported by the Recovery and Resilience Plan (PRR) and by the European Funds Next Generation EU, for the support of this research.
\bibliographystyle{IEEEtran}
\bibliography{IEEEabrv,final}

\newpage
%
%
%
%
%
\section*{Authors}
\begin{IEEEbiographynophoto}{Saulo Queiroz}(sauloqueiroz@utfpr.edu.br, saulo@dei.uc.pt)
is an associate professor at the Federal University of Technology (UTFPR), Brazil, where he has taught courses in algorithms, computational complexity, and signal processing for over a decade. He earned his Ph.D. with highest honors from the University of Coimbra, Portugal. He has contributed to various research projects, including open-source networking initiatives like Google Summer of Code and industry-partnered research. His current research focuses on signal processing algorithms, computational complexity, and wireless networking.
\end{IEEEbiographynophoto}

\begin{IEEEbiographynophoto}{Jo\~ao P. Vilela}(jvilela@fc.up.pt)
is a professor at the University of Porto's Department of Computer Science and a senior researcher at INESC TEC and CISUC. He earned his Ph.D. in Computer Science from the University of Porto in 2011. He has been a visiting researcher at Georgia Tech and MIT, focusing on security topics. His research interests include security and privacy in computer and communication systems, with applications in wireless networks, IoT, and mobile devices. He has coordinated several national and European-funded projects in these areas, covering topics like physical-layer security, next-generation networks, and privacy-preserving data mining.
\end{IEEEbiographynophoto}

\begin{IEEEbiographynophoto}{Edmundo Monteiro}(edmundo@dei.uc.pt)
is a Full Professor at the University of Coimbra, Portugal, from
where he graduated in Electrical Engineering (Informatics Specialty) in 1984. He
has 40 years of research and industry experience in the field of Mobile
Communications, Quality of Service, and Cybersecurity. His publications include
over 200 papers in international refereed journals and conferences. He is member
of the Editorial Board of Springer Wireless Networks and ITU Journal on Future
and Evolving Technologies journals, and a Senior Member of IEEE COMSOC and
ACM SIGCOMM. He is also the Portuguese representative in IFIP TC6
(Communication Systems).
\end{IEEEbiographynophoto}

\vfill
\end{document}